\documentclass{article}

\usepackage{arxiv}

\usepackage[utf8]{inputenc} 
\usepackage[T1]{fontenc}    
\usepackage{hyperref}       
\usepackage{url}            
\usepackage{booktabs}       
\usepackage{amsfonts}       
\usepackage{nicefrac}       
\usepackage{microtype}      
\usepackage{cleveref}       
\usepackage{lipsum}         
\usepackage{graphicx}
\usepackage{natbib}
\usepackage{doi}

\title{Models of attractor dynamics in the brain}

\date{}

\newif\ifuniqueAffiliation
\uniqueAffiliationtrue

\ifuniqueAffiliation 
\author{ \hspace{1mm}Tala Fakhoury* \\
	Center for Theoretical Neuroscience\\
	Columbia University\\
	New York, USA \\
	\texttt{tf2546@columbia.edu} \\
	\And
	\hspace{1mm}Elia Turner* \\
	Department of Mathematics\\
	Technion\\
	Haifa, Israel \\
	\texttt{eliaturner11@gmail.com} \\
      \And 
	\hspace{1mm}Sushrut Thorat* \\
	Institute of Cognitive Science\\
	Osnabrück University\\
	Osnabrück, Germany \\
	\texttt{sthorat@uos.de} \\
    \And
	\hspace{1mm}Athena Akrami \\
	Sainsbury Wellcome Centre\\
	University College London\\
	London, United Kingdom \\
	\texttt{athena.akrami@ucl.ac.uk} \\
}
\else
\usepackage{authblk}

\newbox{\orcid}\sbox{\orcid}{\includegraphics[scale=0.06] 
\author[1]{%
	\href{https://orcid.org/0000-0000-0000-0000}{\usebox{\orcid}\hspace{1mm}David S.~Hippocampus\thanks{\texttt{hippo@cs.cranberry-lemon.edu}}}%
}
\author[1,2]{%
	\href{https://orcid.org/0000-0000-0000-0000}{\usebox{\orcid}\hspace{1mm}Elias D.~Striatum\thanks{\texttt{stariate@ee.mount-sheikh.edu}}}%
}
\affil[1]{Department of Computer Science, Cranberry-Lemon University, Pittsburgh, PA 15213}
\affil[2]{Department of Electrical Engineering, Mount-Sheikh University, Santa Narimana, Levand}
\fi

\renewcommand{\headeright}{}
\renewcommand{\undertitle}{}

\hypersetup{
pdftitle={A template for the arxiv style},
pdfsubject={q-bio.NC, q-bio.QM},
pdfauthor={David S.~Hippocampus, Elias D.~Striatum},
pdfkeywords={First keyword, Second keyword, More},
}

\begin{document}
\maketitle

\begin{abstract}
Attractor dynamics are a fundamental computational motif in neural circuits, supporting diverse cognitive functions through stable, self-sustaining patterns of neural activity. In these lecture notes, we review four key examples that demonstrate how autoassociative neural network models can elucidate the computational mechanisms underlying attractor-based information processing in biological neural systems performing cognitive functions. Drawing on empirical evidence, we explore hippocampal spatial representations, visual classification in the inferotemporal cortex, perceptual adaptation and priming, and working-memory biases shaped by sensory history. Across these domains, attractor network models reveal common computational principles and provide analytical insights into how experience shapes neural activity and behavior. Our synthesis underscores the value of attractor models as powerful tools for probing the neural basis of cognition and behavior.
\end{abstract}


\renewcommand{\thefootnote}{}
\footnotetext{*Equal contribution}
\renewcommand{\thefootnote}{\arabic{footnote}} 

\section*{Introduction} 
The brain's remarkable computational abilities emerge from the complex interplay of neural circuits organized into distinct, yet interconnected, functional architectures. Among these architectural motifs, canonical microcircuits with recurrent connectivity patterns represent fundamental computational units that appear across diverse brain regions~\citep{douglas1989canonical,bastos2012canonical}. These recurrent connections, where neurons form closed loops of excitation and inhibition, are ubiquitous throughout cortical and subcortical structures and play a crucial role in information processing beyond what purely feed-forward architectures could achieve~\citep{van2020going}.

Recurrent neural networks support a class of dynamics known as attractor dynamics, where network activity evolves toward and stabilizes around specific patterns—attractors in the state space of possible neural activations~\citep{hopfield1982neural,amit1989modeling}. These attractors can manifest as point attractors (stable equilibrium states), line or ring attractors (continuous manifolds of stable states), or limit cycles (periodic trajectories), providing the substrate for persistent neural activity essential for working memory, perception, decision-making, spatial navigation, and planning. The computational richness of attractor networks stems from their ability to store multiple stable states, implement pattern completion from partial inputs, resist noise, and support information integration over time~\citep{wang2001synaptic,rolls2007attractor,tang2018recurrent}.


The expressive power of attractor dynamics as models of biological neural computation has received substantial empirical support across a range of cognitive domains. In this review, we summarize two lectures delivered by Dr. Athena Akrami at the 2023 School on Analytical Connectionism—an event focused on analytical tools for probing neural networks and higher-level cognition—held at University College London. The lectures illustrate how attractor-based models can illuminate core neural computations across systems. We begin with bifurcation phenomena in hippocampal place cell remapping, followed by attractor dynamics in inferotemporal cortex during visual classification. We then examine the interplay of adaptation and priming effects in perception, and conclude with coupled attractor dynamics between cortical regions that give rise to working memory biases. Throughout, we highlight how recurrent neural computations and their associated attractor dynamics offer a unifying computational framework for understanding seemingly disparate neural and cognitive phenomena.

\textbf{Broader landscape.} Here we focus on discrete (point) attractors and autoassociative networks because they map cleanly onto the examples covered in the lectures. Attractor theory spans a wider set of influential models. For example, continuous (``bump") attractors have been central for parametric working memory and spatial representations~\citep{compte2000synaptic,wimmer2014bump}. Line-attractor constructions provide a limiting case for stable integration and graded memory~\citep{seung1996brain}. In decision-making, recurrent attractor networks offer a mechanistic account of categorical choices and speed–accuracy trade-offs~\citep{wang2002probabilistic,wong2006recurrent}. These related formulations share the core idea of low-dimensional stable manifolds in population activity, but differ in what is stored (discrete states vs. continuous variables) and which behavioral signatures they are designed to explain.

\section{Attractor dynamics in hippocampal place cells}

\begin{figure}[t]
\begin{center}
\includegraphics[width=0.9\textwidth]{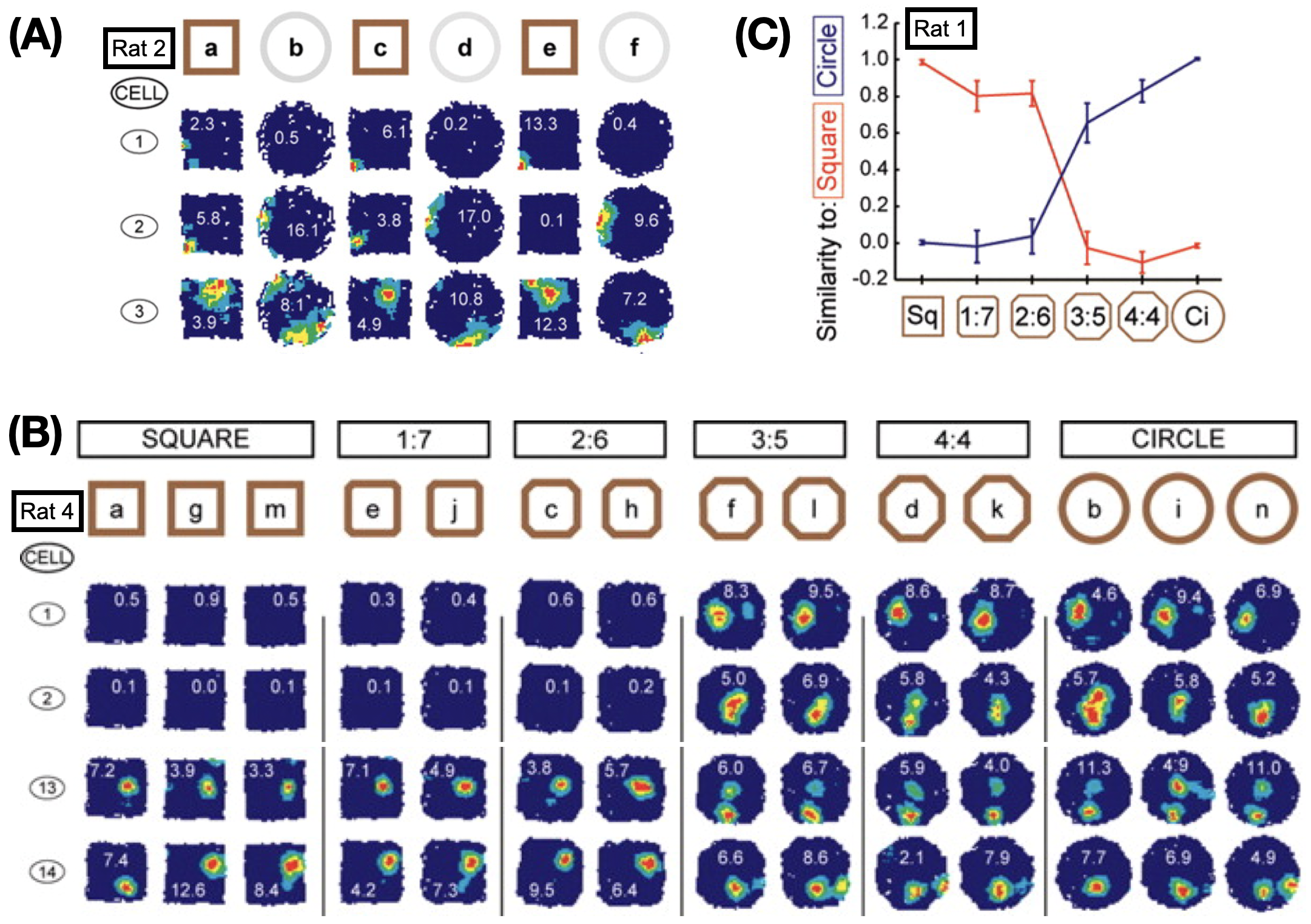}
\end{center}
\caption{Attractor dynamics in hippocampal place cells. \textbf{(A)} Exposure to two distinct ``endpoint" environments led to separate, non-overlapping spatial firing patterns in hippocampal place cells (red is higher, blue is lower). \textbf{(B)} When animals were subsequently exposed to morphed intermediate environments, the place cell firing pattern shifted abruptly to resemble those of the closer endpoint environment (see \citet{wills2005attractor} for details on the measure of similarity of place cells' firing rates between enclosures). \textbf{(C)} Quantification of these transitions confirmed discrete, attractor-like shifts in spatial representations, supporting the existence of attractor dynamics underlying place cell activity. Plots were adapted and reprinted from \citet{wills2005attractor} with permissions.} 
\label{fig:fig1}
\end{figure}

The hippocampus is widely regarded as a hotspot for investigating attractor dynamics, largely due to its central role in encoding spatial representations and memory retrieval~\citep{knierim2012attractor,whittington2020tolman}. In particular, recurrent connectivity within the CA3 region of the hippocampus is believed to implement auto-associative network dynamics, capable of maintaining stable attractor states that support robust spatial representations observed in CA1~\citep{rolls2007attractor}. Hippocampal place cells, which exhibit highly selective firing patterns corresponding to specific locations in space, offer an ideal test-bed for evaluating principles of attractor-based computation and theoretical models grounded in these dynamics.

\citet{wills2005attractor} investigated hippocampal attractor dynamics by examining changes in place cell firing patterns as rats explored environments of varying shapes. Rats were first familiarized over six days with two distinct enclosures, one square and one circular, during which hippocampal place cells in CA1 exhibited global remapping between the two shapes in four out of six animals (Figure~\ref{fig:fig1}A; the remaining two animals either did not show remapping between square and circle enclosures or their remapping did not generalize across enclosures material despite having the same shape). On the seventh day, environments were systematically morphed from square to circle in intermediate (octagonal) shapes (square-like and circle-like morph trials were alternated), and the response of individual place cells was monitored. If the place cell firing patterns are driven by a latent attractor network, then the patterns for the square-like shapes should closely resemble the firing pattern in the square enclosure, and the circle-like shapes should closely resemble the firing pattern in the circular enclosure. Indeed, cells demonstrated an abrupt, attractor-like switch between square-like and circle-like firing patterns (Figure~\ref{fig:fig1}B,C), observable even within the initial $10$ seconds of each morph trial, although these transitions became increasingly pronounced with continued exposure. These findings provide strong empirical support for attractor dynamics in hippocampal spatial representations. Interestingly, contrasting results from \citet{leutgeb2005progressive} showed more graded transitions in similar morphing paradigms, suggesting that the hippocampus can operate in different dynamical regimes—discrete or continuous—depending on factors such as input ambiguity, learning history, and task demands~\citep{tsodyks2005attractor}. Consistent with these observations, computational implementations of hippocampal attractor dynamics in morphing-environment paradigms show that recurrent network dynamics can settle into one of the previously learned spatial maps (often with hysteresis) despite ambiguous intermediate inputs, producing attractor-like transitions between map states~\citep{renno2014signature,romani2010continuous}.

Subsequent studies further refined this view. \citet{colgin2010attractor} demonstrated that abrupt global remapping in the hippocampus cannot be fully accounted for by direct, feature-based differences between environments, as simpler autoassociative models would suggest. Instead, this remapping reflects the activation of distinct path-integration reference frames shaped by environmental features, such as geometry. These reference frames are generated primarily by the medial entorhinal cortex~\citep{hafting2005microstructure}, which provides spatial input to the hippocampus. Together, these results underscore how attractor dynamics within the hippocampal–entorhinal circuit shape spatial memory representations, with environmental geometry and path integration playing critical roles in anchoring these neural cognitive maps.

\section{Attractor dynamics in inferior temporal cortex}

\begin{figure}[t]
\begin{center}
\includegraphics[width=\textwidth]{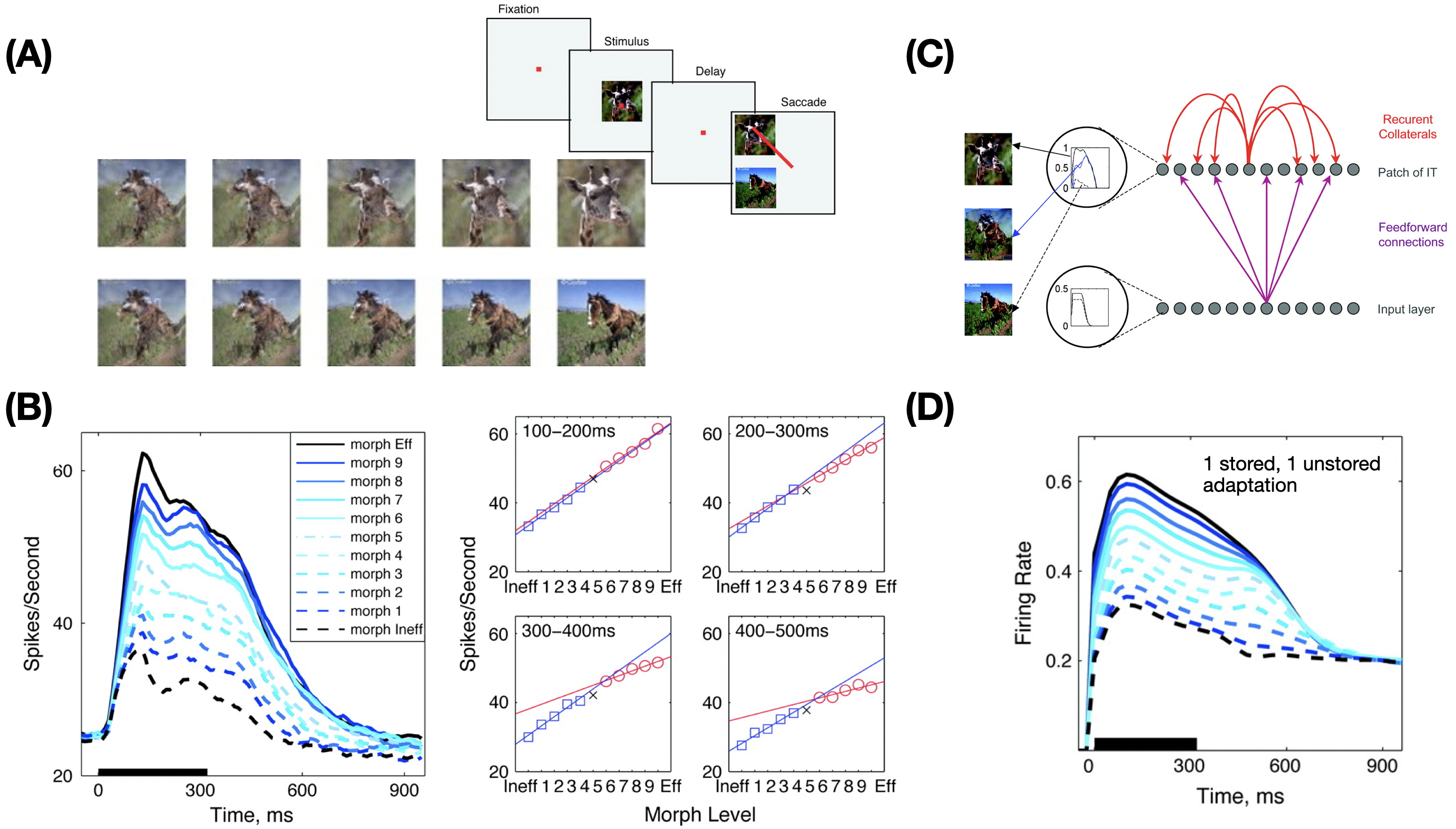}
\end{center}
\caption{Modeling attractor dynamics in inferior temporal (IT) cortex. \textbf{(A)} Two monkeys were trained to perform a match-to-sample task where the sample could be a morph of the two option images. \textbf{(B)} In IT neurons selective to one of the images (``morph eff") but not the other (``morph ineff"), the activity elicited by intermediate morphs closer to the effective image was more similar to the effective image's activity than predicted by a linear dependence on morph level, resembling attractor dynamics ($300\,$ms post stimulus onset and later; solid and dashed lines indicate higher image-based similarity to the effective or ineffective morphs, respectively). \textbf{(C)} An autoassociative neural network model was constructed to simulate these dynamics, with orthogonal patterns stored as memories in the recurrent IT network. As seen in \textbf{(D)}, when memory storage approached network capacity and firing rate adaptation was included, neurons selective for a stored pattern showed similar attractor-like convergence: morphs closer to the memorized pattern elicited neural activity similar to that of the stored pattern, paralleling the experimental observation in (B). Plots were adapted and reprinted from \citet{akrami2009converging} with permissions.} 
\label{fig:fig2}
\end{figure}

The inferior temporal (IT) cortex is critical for visual object recognition~\citep{dicarlo2012does} and long-term visual memory storage~\citep{sakai1991neural}. Given its rich recurrent connectivity and involvement in memory retrieval, IT has been hypothesized to exhibit attractor dynamics similar to those observed in the hippocampus. Attractor network models have been proposed as computational mechanisms to support categorization and stabilize object representations, particularly under conditions of visual ambiguity~\citep{miyashita1993inferior,rolls2007attractor}.

In their study, \citet{akrami2009converging} directly tested this hypothesis, probing attractor-like categorization dynamics in IT using a visual morphing paradigm conceptually analogous to the hippocampal place cell studies discussed earlier. Monkeys performed a match-to-sample task, discriminating between pairs of familiar photographic stimuli (“endpoints”) and intermediate morphs generated via nonlinear pixel-wise blending (Figure~\ref{fig:fig2}A). The match options were always a pair of endpoints, corresponding to the sample, which could be one of the endpoints or their morph. Single-electrode recordings in anterior IT cortex (area TE and adjacent perirhinal cortex) targeted neurons selectively responsive to one of the endpoint images (designated “effective”) but not its paired counterpart (“ineffective”). Early neural responses ($100$-$200\,$ms after stimulus onset) scaled linearly with stimulus similarity to endpoints, encoding the morph level (Figure~\ref{fig:fig2}B). However, during a later response period ($200$–$500\,$ms post-stimulus), firing rates for morph stimuli similar to the effective stimulus showed convergence, losing their linear dependence and approaching the firing rate elicited by the effective endpoint. This convergence was asymmetric: morphs closer to the ineffective endpoint maintained linearly graded responses, suggesting a selective attractor basin biased toward the effective memory. Notably, the strength of this asymmetric convergence grew with the animals' behavioral proficiency, indicating experience-dependent shaping of attractor-like dynamics in IT.

To explore the underlying mechanisms, \citet{akrami2009converging} implemented an autoassociative neural network model (Figure~\ref{fig:fig2}C) comprising two layers of $2500$ neurons each: an input layer mimicking early visual representations and a recurrent output layer simulating the IT cortex. Each output neuron received sparse, randomly assigned feedforward inputs from $750$ neurons in the input layer (modeling neurons in early/mid visual cortex that exhibit sparse activations,~\citet{willmore2011sparse}) and recurrent inputs from $500$ randomly selected neurons within the output layer. 

Visual memory patterns were stored within the recurrent connections using a standard Hebbian covariance learning rule~\citep{rolls1997neural}:
\begin{equation}
    w_{ij} = \frac{1}{C \alpha}\sum_{l=1}^{p} c_{ij} g_i^l (g_j^l - \bar{g})
\end{equation}
where $w_{ij}$ is the synaptic weight between neurons $i$ and $j$, $c_{ij}$ indicates the presence (1) or absence (0) of a connection from neuron $j$ to neuron $i$, $g_i^l$ is the activity of neuron $i$ in pattern $l$, $\bar{g}$ is the mean activity across all patterns, $C$ is the number of recurrent connections per neuron, and $\alpha$ is the activity sparseness parameter.

To simulate memory storage of familiar visual stimuli, a set of random but structured activity patterns (``stored patterns”) was first established. These consisted of sparse firing rate vectors across the $2500$ input-layer neurons, drawn independently from a truncated logarithmic distribution. Each stored pattern corresponded conceptually to one familiar (``endpoint”) visual stimulus used in the experiment. Morph stimuli were then created by systematically blending pairs of stored patterns—analogous to intermediate morphs used in the empirical study—by replacing the firing rate of a randomly selected subset of neurons in one stored pattern with the corresponding values from another. Importantly, the network stored only the endpoint patterns, not the intermediate morphs, consistent with the assumption that monkeys would not form stable memory traces for ambiguous intermediate stimuli that lacked distinct behavioral relevance.

The model included spike-frequency adaptation to mimic firing rate decay observed in cortical neurons. Adaptation was implemented by subtracting a term proportional to each neuron’s recent activity from its total synaptic input, as:
\begin{equation}
    r_i(t) = g \left(h_i(t) - c\left[r_i^1(t) - r_i^2(t)\right] - r_{th} \right)_+
\end{equation}
with:
\begin{equation}
    r_i^1(t) = r_i^1(t-1)e^{-b_1} + r_i(t-1),\quad r_i^2(t) = r_i^2(t-1)e^{-b_2} + r_i(t-1)
\end{equation}

Here, $r_i(t)$ is the activity of neuron $i$ at time $t$, $h_i(t)$ is the summed synaptic input, $r_{th}$ is the firing threshold, $g$ is the neural gain (characterizes the response proportion of a neuron given its input drive), $+$ corresponds to rectified linear activation (ReLU), and parameters $c$, $b_1$, and $b_2$ control the strength and time-scale of adaptation. 

Simulations revealed that asymmetric attractor-like convergence of neural responses emerged specifically when morph stimuli were interpolated between a stored (``effective") and an unstored (“ineffective”) pattern, and the network operated near its maximum memory storage capacity (approximately $160$ patterns in this model; Figure~\ref{fig:fig2}D). Under these conditions, responses to morphs resembling the stored pattern converged toward the corresponding attractor state, while responses to morphs closer to the unstored pattern remained linearly graded (i.e., dependent on morph level). In contrast, when memory load was low (e.g., $20$ patterns), convergence was overly broad with almost all morphs attracted to the stored memory, inconsistent with sharp behavioral categorization and neural responses observed empirically. Furthermore, when both morph endpoints were stored, convergence occurred symmetrically, eliminating the observed asymmetry in IT responses. To explain this, the authors proposed that this discrepancy could reflect the experimental selection bias in neuron sampling: neurons were selected based on their responsiveness to one of the stimuli, likely belonging to the attractor basin of the "effective" pattern, whereas the "ineffective" pattern may have been represented elsewhere in IT~\citep{haxby2001distributed,kiani2007object}. Thus, the observed asymmetry likely reflects this experimental bias in selective sampling of neuronal populations belonging predominantly to one endpoint’s memory representation rather than a fundamental computational difference between stored and unstored patterns. Finally, the inclusion of spike-frequency adaptation improved the model's match to experimental data by allowing network activity to decay over time after stimulus offset and effectively replicating the experimentally observed temporal response dynamics.

Overall, the modeling results support the view that category-specific convergence in IT emerges from attractor dynamics within local recurrent networks, shaped by memory load and stimulus familiarity. This framework provides a mechanistic account of how experience-dependent plasticity gives rise to categorical visual representations in the IT cortex.

\section{Prior experience and perceptual biases}
\label{sec:others}

\begin{figure}[t]
\begin{center}
\includegraphics[width=0.6\textwidth]{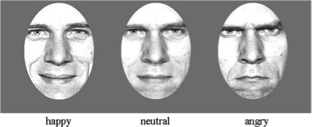}
\end{center}
\caption{Example primer and test stimuli from \citet{webster2004adaptation}, adapted from the JACNeuF and JACFEE image dataset of \citet{biehl1997matsumoto}. The primers "happy" and "angry" (left and right, respectively) are presented before the test stimulus, a "neutral" (center). Behavioral results from the study show that subjects' perception of the neutral face is systematically biased by the primer: participants are more likely to judge the neutral face as "happy" or "angry" depending on whether the preceding primer was happy or angry, respectively.} 
\label{fig:fig3}
\end{figure}

Perception is not a passive reflection of sensory input, but a constructive process profoundly shaped by prior experience that gives rise to systematic perceptual biases~\citep{helmholtz1924treatise,beck1967eye}. When faced with ambiguous stimuli, the brain actively resolves uncertainty by integrating contextual information and drawing on memory-based predictions~\citep{friston2005theory, bar2007proactive}. Prior exposure to prototypical stimuli can bias perception in two seemingly opposing directions: adaptation aftereffects, where perception is repelled away from the recently experienced adapter stimulus, and priming effects, where perception is attracted toward it~\citep{logothetis1995psychophysical, tulving1990priming}. A compelling example occurs in facial emotion perception, where prolonged exposure to a happy face causes a subsequent neutral face to appear slightly angry—a classic repulsive aftereffect~\citep{webster2004adaptation, aguado2007effects}, as shown in Figure~\ref{fig:fig3}. However, a brief exposure to a happy face can cause subsequent faces to be perceived happier—a classic priming effect~\citep{murphy1993affect}.

Several theoretical frameworks have attempted to reconcile these effects by appealing to differences in temporal dynamics, neural substrates, or functional roles. Early formulations, such as adaptation-level theory in \citet{helson1964adaptation}, modeled how prior experiences set perceptual reference points. Later, following Barlow's theory of sensory recalibration and predictive coding frameworks~\citep{barlow1993theory}, these phenomena have been recast within a Bayesian inferential model, in which perception results from the integration of sensory evidence and prior expectations~\citep{kersten2004object,fritsche2017opposite}.

In a different take, \citet{akrami2010attractors} offered a unified mechanistic explanation using attractor neural network models incorporating firing rate adaptation. In this framework, the same visual input can produce either adaptation or priming depending on the temporal dynamics of network activity and the stability of stored memory representations. Specifically, short-lived inputs may nudge activity toward a familiar attractor, yielding priming, while sustained stimulation can destabilize that attractor via adaptation, producing repulsion. This model provides an account of how opposing perceptual biases emerge from a single underlying circuit architecture modulated by memory and experience (see below for details).

\subsection*{From Adaptation to Priming: The Role of Task Delays}

\begin{figure}[t]
\begin{center}
\includegraphics[width=0.8\textwidth]{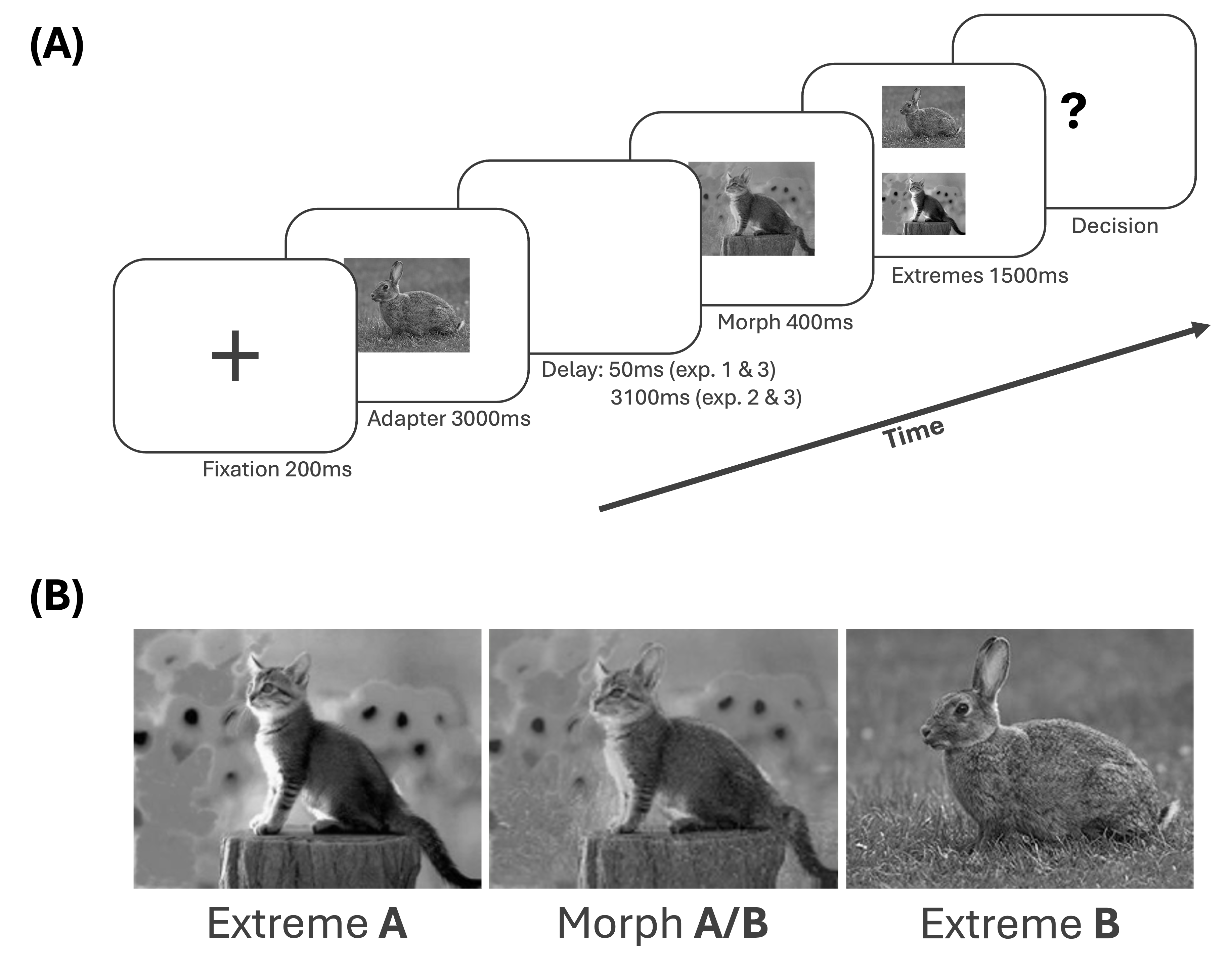}
\end{center}
\caption{Design of the task in \citet{daelli2010recent} probing the influence of primers in perceptual adaptation. In \textbf{(A)}, a primer or "adapter" is first presented for a fixed period of $3\,$s, followed by a delay period of a varied length ($50\,$ms for experiments 1 and 3 and $3100\,$ms for experiments 2 and 3). The morphed object or ambiguous stimulus is then presented for $400\,$ms. The task ends after the subject is probed to answer which stimulus category the morphed image belongs to. \textbf{(B)} Example images shown to the subjects with both extremes (A and B) and the morph (A/B).} 
\label{fig:fig4}
\end{figure}

\begin{figure}[t]
\begin{center}
\includegraphics[width=\textwidth]{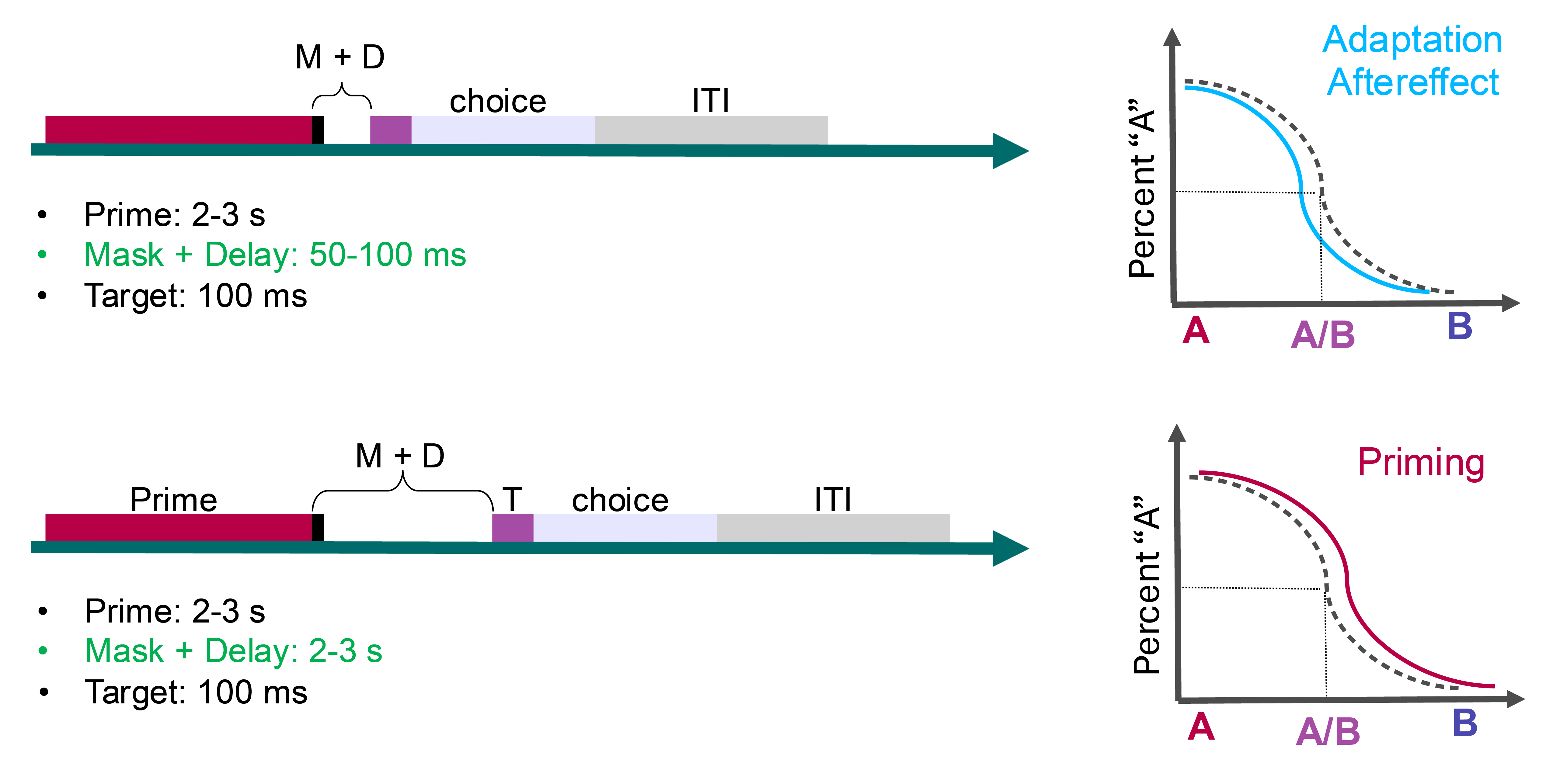}
\end{center}
\caption{Illustrated results from \citet{daelli2010recent} of the priming effects based on delay duration. The top panel shows the experimental paradigm with short delays ($5$-$100\,$ms) between prime and target, resulting in adaptation aftereffects where ambiguous stimuli are perceived as less similar to the adapter (blue curve shows perceptual shift away from prototype A). The bottom panel shows the same paradigm with longer delays ($2$-$3\,$s), where the effect reverses to priming, with ambiguous stimuli perceived as more similar to the adapter (red curve shows perceptual shift toward prototype A). Both conditions used a $100\,$ms target presentation, demonstrating how temporal dynamics determine whether adaptation or priming dominates perception.} 
\label{fig:fig5}
\end{figure}

Adaptation aftereffects represent well-documented phenomena in vision research~\citep{webster1999figural, clifford2007visual}, and while traditionally studied with low-level features such as color and orientation~\citep{gibson1937adaptation,blakemore1969existence}, these effects extend to high-level domains including face~\citep{leopold2001prototype,rhodes2003fitting} and complex object~\citep{daelli2010recent} perception. However, experiments that explore both adaptation and priming effects within a single paradigm, particularly for non-face objects, have been poorly documented.

Building on the attractor-based models of perception and categorization of morphed stimuli discussed in Section 2, \citet{daelli2010recent} investigated how perceptual adaptation influences the interpretation of ambiguous real-world object images and how these perceptual biases evolve. Crucially, their study systematically explored how task parameters-such as the duration of the adapting stimulus, the characteristics of the test image, and the delay interval between the two-modulate perceptual outcomes. 

The authors conducted three behavioral experiments using morphed images of animals, plants, and objects under varying temporal conditions (Figure~\ref{fig:fig4}A). In the first experiment, participants were presented with a clear prototype image (adapter) followed shortly ($50\,$ms delay) by a morph between that prototype and another object. Participants consistently exhibited a repulsive bias, perceiving the morph more dissimilar to the adapter than it actually was when prompted to choose which of the two original images the morph stimulus was closest to (Figure~\ref{fig:fig5}-top), demonstrating adaptation aftereffects in the perception of complex, non-face objects. To establish the temporal stability of these adaptation aftereffects, the second experiment introduced a longer delay between adapter and target ($3100\,$ms). Surprisingly, the repulsive effect disappeared and was replaced by an attractive priming effect—participants were now more likely to judge the ambiguous morph as resembling the adapter (Figure~\ref{fig:fig5}-bottom). This temporal reversal suggests that adaptation effects weaken over time, revealing a slower-developing priming mechanism that biases perception toward previously seen stimuli. In the final experiment, the researchers tested whether adaptation could still occur when the adapter itself was ambiguous. Unlike in previous experiments, adaptation to an ambiguous stimulus consistently led to priming, regardless of the duration of the delay. This finding suggests that adaptation requires a well-defined, strongly encoded prototype to exert a repulsive influence on perception, while ambiguous stimuli are more likely to produce attractive biases.

Collectively, these findings demonstrate that the same stimuli and task can elicit either adaptation or priming effects, depending on temporal task parameters, highlighting the dynamic nature of perceptual biases. The experiments reveal two opposing yet interrelated effects: adaptation, which can induce repulsive aftereffects by attenuating responses to recently encountered stimuli, and memory-based priming, which attracts perception toward prior stimuli via the activation of stable memory representations. When a stimulus is well-defined and recently presented, adaptation dominates, pushing perception away. As time passes, or when adaptation is weak (as with ambiguous stimuli), priming emerges, pulling perception toward familiar experiences. Together, these results  provide a clear empirical characterization of how perceptual history modulates ongoing experience and set the stage for the following section, where we show these effects can be captured mechanistically within a unified attractor network framework, governed by the balance between adaptation and memory-based priming.

\subsection*{A unifying attractor dynamics model to short-term visual experience}

\begin{figure}[t]
\begin{center}
\includegraphics[width=0.8\textwidth]{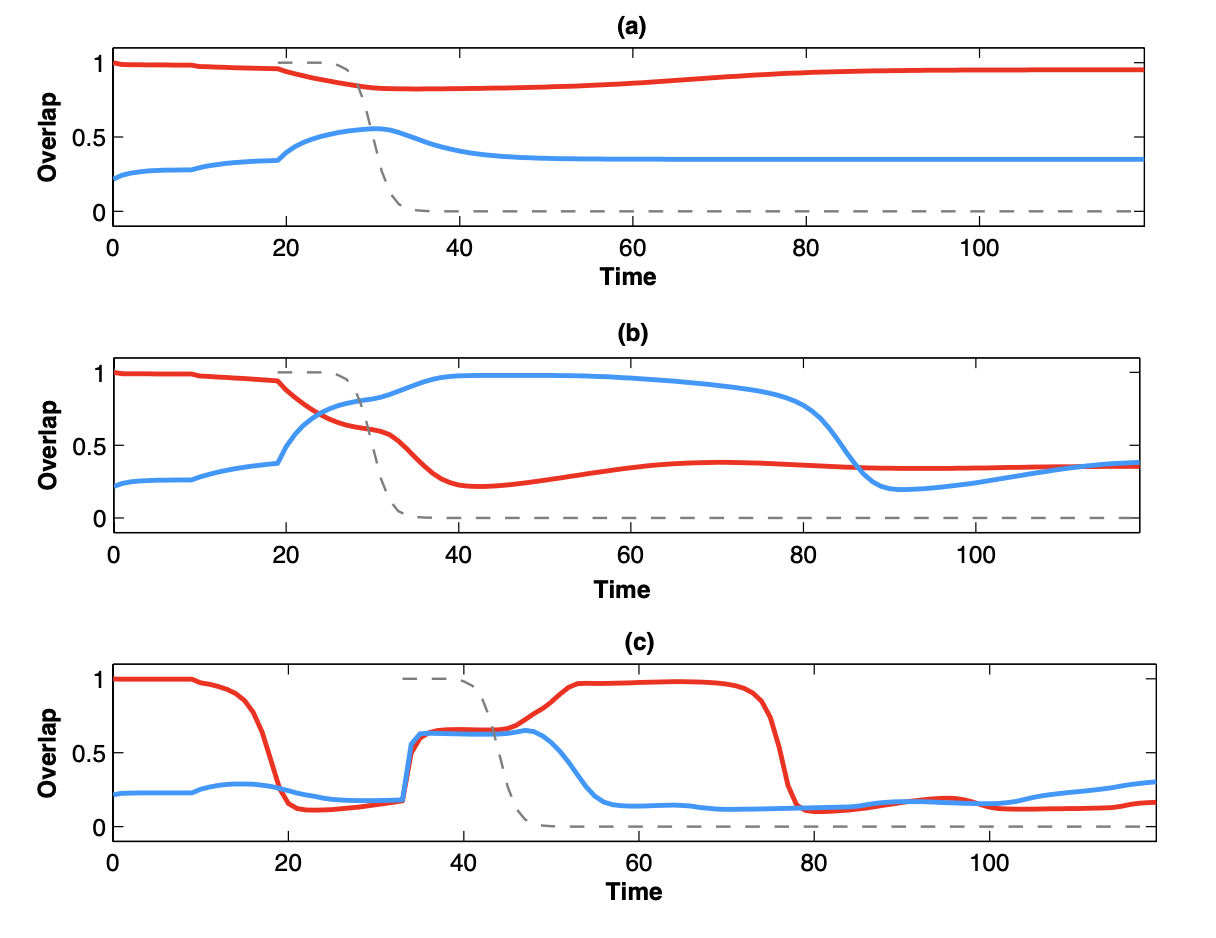}
\end{center}
\caption{Comparing perceptual effects in an attractor neural network model. The figure illustrates different response patterns in the network. The red line represents the differential overlap with pattern A, ${m^{A}_{A}(t) - m^{A}_{B}(t)}$ (i.e. when A is the adapter versus when B is the adapter), while the blue line shows the corresponding differential overlap for pattern B, ${m^{B}_{A}(t) - m^{B}_{B}(t)}$. The top panel demonstrates 'Type A priming' resulting from perceptual bias mechanisms, the middle panel shows adaptation aftereffects with opposing directional shifts, and the bottom panel illustrates 'Type B priming' where extended delay intervals produce double adaptation aftereffects, effectively resulting in a priming effect. Reproduced from \citet{akrami2010attractors}.} 
\label{fig:fig6}
\end{figure}

The studies reviewed above demonstrate that perception is inherently contextual, requiring the integration of prior experience with incoming sensory input to resolve categorical decisions. This process is fundamentally rooted in the temporal dynamics of the underlying neural circuitry. To investigate this, \citet{akrami2010attractors} sought to bridge the gap between historically disparate perceptual phenomena—adaptation aftereffects and priming—by developing a unified neural network model grounded in attractor dynamics. The authors' primary intention was to demonstrate that these opposing perceptual biases do not require separate mechanisms but instead emerge from the intrinsic interaction between recurrent excitation and firing rate adaptation. By situating these effects within a single biophysical framework, their model successfully reproduces key electrophysiological and behavioral findings from \citet{daelli2010recent} and provides a mechanistic account for the diverse range of priming effects observed in humans.

The network architecture is an auto-associative memory model comprising recurrently connected neurons with threshold-linear activation functions. Crucially, it incorporates firing rate adaptation, a biophysical property of pyramidal neurons, which allows the network to transition between transient and stable states. The investigation centers on how the interplay between recent input-driven activity and stable attractor dynamics gives rise to systematic perceptual biases.

To quantify these effects in the model given different stimuli and under different experimental settings, the authors define a similarity index measuring the distance between two network states at any time:

\begin{equation}
    R = m_{A}^{A}(t) - m_{A}^{B}(t)
\end{equation}

where $m^{a}_{a}(t)$ is the overlap with pattern A when A is the adapter, and $m^{a}_{b}(t)$ is the overlap with A when B is the adapter. This metric allowed the authors to systematically map the influence of adaptation strength, delay, and target duration on perceptual outcomes across a broad range of parameter space and as a function of time in a given trial.

The simulations reveal that firing rate adaptation (FRA) is the essential component for producing adaptation aftereffects, acting as a time-dependent negative feedback mechanism that modulates the stability of attractor states. By systematically varying parameters—including memory load, FRA strength, and stimulus ambiguity—the model maps how network dynamics shift between distinct perceptual regimes (Figure \ref{fig:fig6}).

\begin{itemize}
    \item \textbf{Perceptual Bias} (Type A Priming): In regimes where adaptation is minimal or absent, dynamics are dominated by recurrent collateral excitation. Synaptic reverberation allows the network to remain in the induced attractor state even after the adapter is removed. Consequently, the network remains biased toward the original attractor upon target presentation (Figure \ref{fig:fig6}a).
    
    \item \textbf{Adaptation Aftereffects} (Negative Bias): Under strong firing rate adaptation, the sustained firing of neurons representing the adapter pattern builds up an adaptation current ($I_a$). This increases the firing threshold for those specific neurons, effectively destabilizing the active attractor. When an ambiguous target is presented, the network is ``repelled'' from the fatigued state and settles into a competing attractor (Figure \ref{fig:fig6}b).
    
    \item \textbf{Double Adaptation} (Type B Priming): Under stronger firing rate adaptation, the network may undergo multiple transitions between attractor states. If the initial adaptation is strong enough to push the network out of the first attractor during a delay period, specific temporal conditions allow the network to settle back into the original attractor upon target presentation, paradoxically producing priming (Figure \ref{fig:fig6}c).
\end{itemize}

The research also examined backward masking effects, where a secondary stimulus disrupts initial processing. If a mask is presented before the network settles into an attractor, the build-up of the adaptation current is insufficient to destabilize the state, substantially weakening aftereffects and leaving predominantly priming effects. Furthermore, the temporal relationship between stimuli proved critical. Brief adapter-target intervals favored adaptation aftereffects due to peak $I_a$ levels, while extended delays or longer target exposures promoted priming as the adaptation current decayed, allowing stable attractor representations to regain dominance. 

By integrating these diverse factors into a unified framework, the model demonstrates that the complex interplay between recurrent excitation and firing rate adaptation provides a robust neural mechanism capable of reproducing the full spectrum of observed behaviors. It reveals that shifting experimental conditions—such as stimulus timing, masking, or ambiguity—simply move the system across different operational regimes of the same underlying attractor landscape. Ultimately, these findings underscore the importance of firing rate adaptation as a key modulator of perceptual dynamics and highlight how attractor networks provide a mechanistic basis for the flexible, experience-dependent biases observed in human perception.

\section{Coupled PPC and PFC Dynamics Underlying Biases in Working Memory}
\label{sec:ppc_pfc_wm}

\begin{figure}[ht!]
    \centering
    \includegraphics[width=0.9\textwidth]{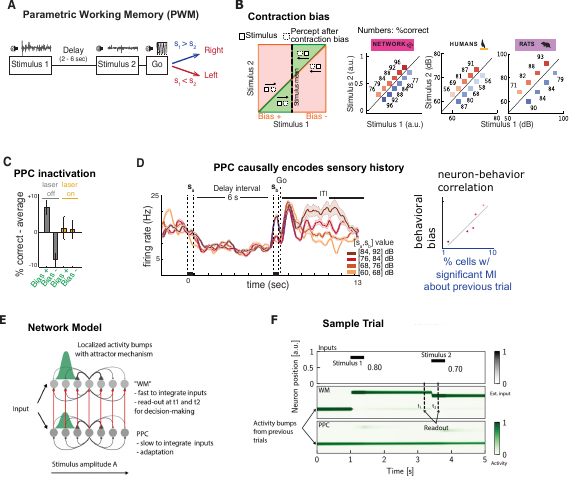}
    \caption{
\textbf{(A)} Auditory parametric working memory (PWM) task from \citet{akrami2018}. Subjects compare two sequentially presented stimuli (\( s_1 \), \( s_2 \)) separated by a delay interval and report which stimulus is louder.
\textbf{(B)} Behavioral contraction bias: the internal memory representation of \( s_1 \) is biased toward the mean of previously experienced stimuli, improving (bias$+$) or impairing (bias$-$) discrimination depending on stimulus order. This effect is observed in humans, rats, and recurrent network models when performance is evaluated for stimulus pairs symmetrically positioned around the diagonal.
\textbf{(C)} Causal evidence for PPC involvement: optogenetic inactivation of PPC during the delay abolishes contraction bias without impairing overall task performance.
\textbf{(D)} PPC neurons encode sensory history across trials. Neural activity during the delay reflects information about stimuli presented in previous trials, and neuron–behavior correlations increase with the strength of history encoding.
\textbf{(E)} Computational model from \citet{boboeva2024}. The model consists of two coupled one-dimensional continuous attractor networks: a fast working-memory (WM) network that encodes the current stimulus, and a slow PPC network that integrates sensory inputs across trials via neuronal adaptation. The PPC provides a history-dependent input to WM but does not directly encode the current stimulus.
\textbf{(F)} Model dynamics during a representative correct trial. Presentation of \( s_1 \) and \( s_2 \) rapidly shifts the activity bump in the WM network, determining the decision. In contrast, the PPC bump remains relatively stable within a trial, reflecting accumulated sensory history rather than the current stimulus. This asymmetric timescale separation allows PPC activity to bias WM representations toward the stimulus mean across trials, reproducing contraction bias while preserving trial-level discrimination.
Adapted from \citet{boboeva2024} and \citet{akrami2018}.
}
    \label{fig:contraction_bias}
\end{figure}

Behavior in perceptual and working memory tasks is often shaped by past sensory experiences. Two prominent forms of such biases are the \textit{contraction bias}- where working memories of stimuli are pulled toward the long-term average of past inputs~\citep{Jazayeri2010, Raviv2012}-and the \textit{recency bias}, or \textit{serial dependence}, where recent stimuli exert a disproportionate influence on current judgments~\citep{fischer2014, barbosa2020}. Although traditionally viewed as distinct phenomena arising from different cognitive mechanisms, recent evidence suggests they may share a common neural substrate. Notably, \citet{akrami2018} showed that silencing the posterior parietal cortex (PPC), in rats, significantly reduces both contraction and recency biases, implicating this region in integrating sensory history. Building on these findings, \citet{boboeva2024} proposed a mechanistic model showing how interaction between the PPC and a downstream working memory (WM) area (e.g., prefrontal cortex, though not proven yet) could give rise to these working memory biases. In what follows, we examine these two studies in detail, focusing on how PPC-WM dynamics may underlie sensory-history-dependent distortions in working memory.

To explore these biases experimentally, \citet{akrami2018} utilized a parametric working memory (PWM) task, involving sequential presentation of two graded stimuli separated by a delay interval. In their auditory version of the task (illustrated in \autoref{fig:contraction_bias}A), rats are presented sequentially with two tones, \(s_a\) and \(s_b\), and must determine which tone was louder, thereby engaging working memory processes rather than immediate sensory comparisons. This delay interval provides a window for prior sensory experience to bias the internal representation of \(s_a\).

A key aspect of the PWM task is the systematic manipulation of stimulus pairs (presented at each trial) based on their sensory values and proximity to the identity line \(s_1 = s_2\) (\autoref{fig:contraction_bias}B). Pairs closer to the diagonal line differ less in intensity. Rats were presented with stimulus pairs that were all equally distant from the diagonal (see \autoref{fig:contraction_bias}B). Despite this uniform structure, performance was systematically modulated by contraction bias: the remembered value of \(s_1\) appeared to be shifted toward the average of previously encountered stimuli (see \autoref{fig:contraction_bias}A). When this shift increased the perceived distance from \(s_2\) (bias+), performance improved; when it reduced the difference (bias-), performance declined.

To quantify these history effects, \citet{akrami2018} used a logistic regression to model behavioral choices as a function of current stimulus values \( (s_1, s_2) \), the average of past stimuli (capturing contraction bias), the stimulus pair from the immediately preceding trial (capturing recency), and reward and choice history. This analysis revealed clear evidence for both contraction and recency biases. Importantly, similar biases were observed in human participants, suggesting conserved underlying mechanisms.

To establish a causal role for PPC, optogenetic inactivation experiments were conducted. Temporary silencing of PPC during the delay period abolished both biases (\autoref{fig:contraction_bias}C), without impairing overall task performance, indicating that PPC specifically contributes to integrating sensory history into working memory representations rather than maintaining WM information per se. Additionally, the PPC was shown to encode sensory history (\autoref{fig:contraction_bias}D).

Building on this, \citet{boboeva2024} developed a computational model comprising two interacting line-attractor networks (\autoref{fig:contraction_bias}E): a slowly integrating PPC network with adaptive dynamics and a WM network downstream to PPC  with more stable, persistent activity that responds to sensory inputs with fast dynamics. Both networks receive direct sensory input, with the PPC accumulating sensory information over time and projecting to the WM, thereby modulating its memory representations.

Sensory-history biases naturally emerge from differences in integration timescales between the PPC and WM networks. During the delay interval, the memory representation in the WM network either remains stable or is shifted by PPC input, introducing bias into working memory. See \autoref{fig:contraction_bias}F for a sample trial. Crucially, the model predicts stronger biases with increased delay intervals, aligning with experimental observations. Moreover, when PPC input was removed from the model, the biases disappeared, mirroring the inactivation findings from \citet{akrami2018} and reinforcing the PPC’s central role as a sensory history integrator.

To reveal the essential statistical mechanism underlying contraction bias in the attractor network, \citet{boboeva2024} also formulated a simplified probabilistic model that abstracts the network dynamics. In this framework, the comparison between the remembered first stimulus $s_1$ and the second stimulus $s_2$ is treated probabilistically: on a fraction of trials, the memory of $s_1$ remains veridical, while on others it is replaced by a sample drawn from the distribution of PPC bump locations, reflecting accumulated sensory history. Behavioral errors arise when this sampled value leads to an incorrect comparison with $s_2$, and the probability of such errors can be computed analytically as integrals over the sampling distribution. Fitting this model to both the network output and empirical behavior of rats and humans shows that it captures key signatures of contraction bias with only a single free parameter, demonstrating that history-dependent biases in the $s_2 - s_1$
 comparison can emerge as a statistical consequence of PPC-driven sampling rather than gradual drift of the memory trace. This probabilistic abstraction thus provides a compact, predictive link between mechanistic network dynamics and observed behavior, and can be related to normative Bayesian descriptions under specific assumptions.

Together, these studies underscore a critical role for PPC-WM interactions in generating working memory biases driven by sensory history. They provide a compelling framework linking behavioral phenomena such as contraction and recency biases to specific neural mechanisms, integrating experimental, computational, and theoretical perspectives on memory distortions.

\section*{Conclusion}

Taken together, the four examples reviewed in these lectures highlight the utility of auto-associative neural networks as a powerful and versatile class of analytical models for studying attractor dynamics—a computational motif that recurs across diverse brain regions and cognitive domains. While these models are intentionally simplified and may not capture the full complexity of neural responses to naturalistic stimuli or real-world behavioral tasks, their strength lies in their analytical tractability. By enabling researchers to precisely isolate, manipulate, and interpret specific features of neural dynamics, these models yield mechanistic insights that are often obscured in more complex, high-dimensional systems. Importantly, auto-associative networks provide a conceptually transparent bridge between neural activity and behavior, revealing how stable patterns of activity can support memory, perception, categorization, and decision making. These insights complement the growing class of large-scale, image-computable neuroconnectionist models that prioritize biological realism and scale~\citep{doerig2023neuroconnectionist}.

While future work must address how to extend attractor-based frameworks to operate in more realistic, high-dimensional sensory spaces~\citep{goetschalckx2023computing,thorat2023characterising,soo2024recurrent}, the foundational principles uncovered by analytical models remain essential. Their simplicity is not a limitation but a strength, offering a conceptual framework for hypothesis testing, theory building, and ultimately, for understanding the computational architecture of the brain. 

\textbf{Assumptions, limitations, and complementary frameworks.} These analytical attractor models make deliberate abstractions (e.g., rate-based units, stylized inputs, simplified noise/adaptation), so they should be read as mechanistic ``existence proofs" rather than complete accounts in naturalistic settings. Importantly, ``alternative" frameworks often target different levels of description rather than different mechanisms: for example, drift–diffusion models~\citep{bogacz2006physics} summarize choice/RT statistics and can be interpreted as a low-dimensional approximation to recurrent decision dynamics~\citep{wang2002probabilistic,wong2006recurrent}, while predictive-processing/predictive-coding views emphasize how recurrent feedback can implement inference via prediction-error correction~\citep{barlow1993theory,friston2005theory}. In this sense, these perspectives are typically complementary to—not in competition with—attractor-based circuit mechanisms.










\bibliographystyle{plainnat}
\bibliography{references}  

\end{document}